\documentclass{egpubl}
\usepackage{eurovis2026}

\SpecialIssueSubmission    

\CGFccby

\usepackage[T1]{fontenc}
\usepackage{dfadobe}  
\usepackage{amsmath}
\usepackage{xspace}
\usepackage{cite}  
\usepackage[ruled,vlined]{algorithm2e}
\usepackage[inkscapelatex=false]{svg} 
\usepackage{booktabs}
\usepackage{multicol}
\usepackage{multirow}
\usepackage{pifont}
\newcommand{\cmark}{\ding{51}}%
\newcommand{\xmark}{\ding{55}}%
\usepackage{subcaption}
\usepackage{float}
\usepackage{caption}

\BibtexOrBiblatex
\electronicVersion
\PrintedOrElectronic

\usepackage{svg}
\usepackage{egweblnk}
\usepackage[most]{tcolorbox}
\definecolor{lightergray}{HTML}{dddddd}
\newtcolorbox{simplebox}{
  colback=white,        
  colframe=lightergray,        
  boxrule=2pt,        
  arc=2mm,              
  enhanced,             
  left=0pt,             
  right=0pt,
  top=0pt,
  bottom=0pt,
  hbox
}

\newcommand{\etal}{\emph{et al.}\xspace}

\title[ARCOL]%
      {ARCOL: Aspect Ratio Constrained Orthogonal Layout}

\author[Z. Alsuwaykit \etal]
{\parbox{\textwidth}{\centering Zainab Alsuwaykit$^{1}$\orcid{0009-0009-7444-445X}
        , Yousef Rajeh$^{1}$\orcid{0009-0004-9001-789X}, Alexandre Kouyoumdjian$^{1}$\orcid{0009-0005-0915-369X}, Steve Kieffer$^{2}$\orcid{0009-0002-7038-9237}, Dominik Engel$^{1}$\orcid{0000-0002-5766-7215}, Sara Di Bartolomeo$^{3}$\orcid{0000-0001-9517-3526}, Martin N\"ollenburg$^3$\orcid{orcid.org/0000-0003-0454-3937} and Ivan Viola$^{1}$\orcid{0000-0001-5923-423X} 
        }
        \\
{\parbox{\textwidth}{\centering $^1$King Abdullah University of Science and Technology, Saudi Arabia\\
$^2$ Alpine Mathematics, Saratoga Springs NY, USA.
$^3$ Technische Universit\"at Wien, Austria
       }
}
}
\teaser{
\centering
\includegraphics[width=\textwidth]{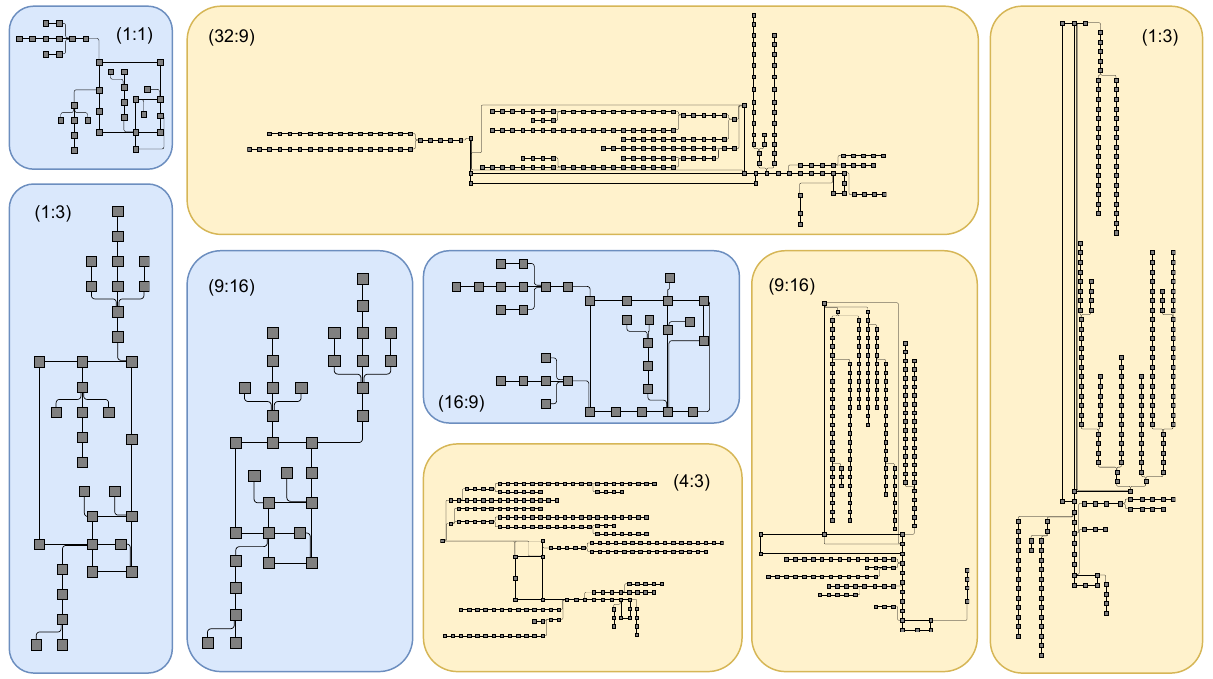}
\vspace{-6mm}
\caption{
   Two different graphs are laid out in different aspect ratios. Blue: small graph from Rome dataset. Yellow: Melbourne metro map.
  }
\vspace{3mm}
}

\begin{document}


\maketitle
\begin{abstract} 
Orthogonal graph layout algorithms aim to produce clear, compact, and readable network diagrams by arranging nodes and edges along horizontal and vertical lines, while minimizing bends and crossings. Most existing orthogonal layout methods focus primarily on quality criteria such as area usage, total edge length, and bend minimization. Explicitly controlling the global aspect ratio (AR) of the resulting layout is as of now unexplored. Existing orthogonal layout methods offer no control over the resulting AR and their rigid geometric constraints make adaptation of finished layouts difficult. With the increasing variety of aspect ratios encountered in daily life, from wide monitors to tall mobile devices or fixed-size interface panels, there is a clear need for aspect ratio control in orthogonal layout methods. To tackle this issue, we introduce Aspect Ratio-Constrained Orthogonal Layout (ARCOL). Building upon the Human-like Orthogonal Layout Algorithm (HOLA)~\cite{Kieffer2016}, we integrate aspect ratio at two different stages: (1) into the stress minimization phase, as a soft constraint, allowing the layout algorithm to gently guide node positions toward a specified target AR, while preserving visual clarity and topological faithfulness; and (2) into the tree reattachment phase, where we modify the cost function to favor placements that improve the AR. We evaluate our approach through quantitative evaluation and a user study, as well as expert interviews. Our evaluations show that ARCOL produces balanced and space efficient orthogonal layouts across diverse aspect ratios.

\ccsdesc[300]{Visualization Techniques~ Graph drawing}
\ccsdesc[300]{Theory of computation~ Design and analysis of algorithms}
\printccsdesc{}%
\end{abstract} 
\newpage

\section{Introduction}

Graphs are a common way to model and show relationships between entities. Node-link diagrams are widely used to make sense of complex systems because they provide an intuitive depiction of connectivity~\cite{Ware2012,Kieffer2016}. Next to straight-line node-link diagrams, orthogonal layouts, where edges run along horizontal and vertical lines, are among the most popular drawing styles. Standards in electrical and software engineering, process flow diagrams, and life‑science notations like SBGN and KGML use orthogonal drawings~\cite{Novere2009,Batini1986}. Their grid-like appearance makes them easy to understand, and they are the default choice for circuit diagrams, UML class diagrams, and database schemas. Many professional tools therefore provide orthogonal layout functionality, and researchers have developed algorithms to help produce clear and compact drawings~\cite{dg-popda-13,yWorks}.

Most existing orthogonal layout algorithms optimize clarity and compactness. They align nodes on a grid and aim to reduce edge crossings and bends, shorten edge lengths and use the drawing area efficiently~\cite{Papakostas1998,Foessmeier1997,bk-bhogd-98,HELEN2002}. However, these methods lack a way to control the global aspect ratio of the graph. Without control over how wide or tall a graph becomes, orthogonal layouts may fit poorly into the available screen space on wide monitors or tall mobile devices. The lack of aspect‑ratio control has become more problematic as visualizations are deployed on many kinds of screens and within fixed‑size interface panels. In biological pathway visualization, for example, modern systems like Metabopolis~\cite{Wu2019} decompose large pathways into sub‑graphs and apply orthogonal layouts, where these sub‑graphs may need to fill predefined rectangular panels with very different proportions. In addition, AR control is becoming important in dashboard-style visualizations, where the screen is divided into many small components with fixed sizes. Each component must use its limited space efficiently, and a graph that does not match the assigned aspect ratio either wastes space or becomes squeezed. Layouts that can adapt to a required aspect ratio make dashboards more readable and more space-efficient. Another use case comes from ultra-wide displays, such as the very wide 360-degree screen used in a recent graph-drawing contest~\cite{dibartolomeo_et_al:LIPIcs.GD.2025.41}. These displays are challenging for traditional layout algorithms because they stretch far beyond standard proportions. Graph layouts that can intentionally adapt to extreme aspect ratios are easier to view and compare on such screens.

The requirement to match a given aspect ratio arises in many application domains, but is currently addressed only in specialized layouts, for instance, with balanced aspect‑ratio trees for straight‑line drawings~\cite{Duncan1998,Junger2018}. Very few efforts have been made to adapt layered drawing approaches to control the graph area or width. For example, R{\"u}egg \etal~\cite{Ruegg2011,RueggEtal2017} explore layering heuristics under width constraints.
Force‑directed layouts can incorporate boundary forces or geometric constraints~\cite{Zhang2016};  however, these techniques do not translate directly to the global aspect ratio of the graph or the orthogonal layouts because orthogonal layouts have rigid geometric constraints and a distinct multi‑phase pipeline. Within the orthogonal domain, aspect ratio remains unexplored, and our interviews with experts of the field confirmed that no current algorithm directly optimizes aspect ratio inside the layout process (see Sec.~\ref{sec:interviews}).

To address this gap, we present ARCOL, Aspect Ratio Constrained Orthogonal Layout. ARCOL extends the state‑of‑the‑art HOLA~\cite{Kieffer2016}, Human‑like Orthogonal network Layout Algorithm and introduces a soft aspect ratio constraint into two phases of the layout pipeline. First, during stress minimization of the core after removing the trees, we add a normalization step that gently encourages the node positions to approach a target aspect ratio (e.g., 1:1, 16:9 or 9:16), while preserving the graph’s readability and quality. Second, in the tree re-attachment phase we introduce a cost that encourages  placement of trees which improves the overall aspect ratio. The special aspect of ARCOL is that it integrates aspect ratio awareness throughout the optimization rather than post-processing it, thus it can gradually adjust the layout without creating unnecessary edge bends or disrupting the graph’s structure. The resulting graphs retain the clarity, symmetry and compactness of HOLA while fitting neatly into frames of different aspect ratios.

We evaluate ARCOL on graphs from the Rome dateset~\cite{DiBattista1997} and compare our results across a wide range of aspect ratios. Through extensive quantitative and qualitative evaluation, we demonstrate that our approach produces balanced layouts and uses screen space efficiently. We employ universal graph metrics~\cite{Mooney2025}, a user study and expert interviews to support these findings. Our study participants and experts show significant preference for diagrams produced by ARCOL and confirm that ARCOL produces reasonable and consistent diagrams across aspect ratio targets. Further, experts note that integrating aspect ratio control into the layout algorithm solves a long‑standing complaint about diagrams being too wide or too tall. 

Our objective is to provide a method with direct control over the aspect ratio, able to fit modern display environments without compromising quality. We will make the code of our method publicly available upon acceptance on (\httpAddr{//osf.io/z8qtd/overview?view_only=30648ded016346aca418cd1e31c28453}).
In summary, our contributions are threefold:
\begin{description}
    \item[Aspect ratio‑aware orthogonal layout:] We propose ARCOL, the first orthogonal layout algorithm that optimizes for a specified aspect ratio of the resulting diagram.
    \item[Integration with HOLA:] We extend the HOLA framework with a soft aspect ratio constraint in the stress minimization phase and a tree reattachment cost, enabling gradual adjustment of node positions without sacrificing clarity.
    \item[Comprehensive evaluation:] We demonstrate the effectiveness of ARCOL through extensive evaluation, including metrics, a user study and expert feedback, showing that ARCOL produces balanced, readable diagrams across aspect ratios and graphs.
\end{description}



\section{Related Work}

Over the years, many orthogonal layout algorithms have focused on bend minimization, compactness, and readability. Another important class of layout algorithms that ARCOL uses for the soft geometric adjustment step is force-directed algorithms, which position nodes using geometric forces to achieve visually balanced layouts. However, despite these advances, the global aspect ratio of the graphs has remained an implicit outcome rather than a controllable parameter. The following overview summarizes major developments in orthogonal and force-directed layouts research.

\subsection{Orthogonal Graph Layouts}


Research on orthogonal graph layout has long recognized the importance of area efficiency and aspect ratio as key aesthetic and functional metrics. Early studies on orthogonal graph drawing primarily focused on minimizing area and bends while maintaining readability and structural clarity. An early foundational work~\cite{Papakostas1998} established a linear time algorithm for constructing orthogonal layouts with small area and few bends. Their approach aimed to balance width and height within nearly square bounds. Complementary to this line of research, Fößmeier and Kaufmann~\cite{Foessmeier1997} addressed high-degree and nonplanar graphs using the Kandinsky and GIOTTO models, which allow rectangular vertex representations and controlled vertex scaling to manage degree constraints. Building upon this foundation, Biedl and Kaufmann~\cite{Biedl1997} advanced the study of area efficiency by proposing an algorithm that achieved near-optimal grid size for both static and incremental orthogonal layouts. Although their constructions often lead to reasonably balanced layouts, aspect ratio is not treated as an explicit optimization objective. 
These aforementioned works improve area bounds but do not offer user-directed control over the graph’s global aspect ratio. Subsequent work incorporated user constraints and stylistic objectives into orthogonal frameworks. Brandes \etal~\cite{Brandes2002} proposed Sketch-Driven Orthogonal Layouts, extending the Kandinsky model with dynamic adjustment to preserve a given sketch's appearance while minimizing bends. Their system followed the topology-shape-metrics (TSM) framework and demonstrated that orthogonal compaction can be integrated with user-guided or dynamic inputs. Their work focuses on orthogonal compaction and does not incorporate any aspect ratio constraint. More recent approaches revisit aspect ratio within compact and grid-structured visualizations. 



Trees have received particular attention in aspect ratio studies because their hierarchical and recursive structure makes AR control tractable and interpretable. Balanced Aspect Ratio Trees were introduced by Duncan \etal~\cite{Duncan1998}, proposing a method for recursively balancing subtrees to maintain bounded aspect ratios while minimizing empty space. Chan \etal~\cite{Chan2002} optimized area and aspect ratio in straight-line orthogonal layouts of binary trees, demonstrating that near-optimal area can be achieved while allowing user-defined aspect ratio control. Felsner \etal~\cite{Felsner2022} generalized the problem to rectangular subdivisions by defining aspect ratio universal layouts, where the assigned aspect ratios of rectangles can be realized without altering adjacency structure. These approaches rely on tree structure or rectangular subdivisions but are not applicable to arbitrary graphs, and do not provide a global AR constraint for general orthogonal layouts.


While prior research has extensively explored area efficiency, bend minimization, and node-level aspect ratio control, none of the existing methods explicitly treat the aspect ratio (the ratio of the graph's width to its height) as a global layout constraint. Ideally, one wants to fit the graph within a predefined display window or screen size without distorting its structure. Early work gave little attention to explicit aspect ratio constraints, but more recent research addresses them. Nachmanson \etal~\cite{Nachmanson2008} developed a heuristic for creating a graph layout with a given aspect ratio within the Sugiyama framework for layered, hierarchical graphs. Essentially, they adjust the number of layers or the distribution of nodes per layer to utilize a given rectangular area well. Jünger~\etal~\cite{Junger2018} present a flow-based horizontal coordinate assignment that supports a prescribed width for the graph. By treating the maximum width as a constraint in the optimization, they produced layered graphs that fit within a fixed width. However, they are only valid for layered graphs, not for orthogonal layouts and the aspect ratio control is achieved through layer width/height adjustments, not geometric optimization. When drawing disconnected graphs or multiple components, one can treat it as a packing problem to meet an aspect ratio constraint. Freivalds \etal~\cite{Freivalds2002} proposed a polynomial packing method for disconnected graph layout. They compute compact shapes for each component and then pack these like rectangles into a target container. Their approach can be seen as trying to minimize unused space for a given bounding rectangle which is effectively a step towards enforcing aspect ratio. Additional work on schematic layouts has explored aspect ratio restrictions as part of preserving the mental map~\cite{Takahashi2019}. Their approach imposes constraints to limit the variation of regional aspect ratios within the railway map, helping preserve the overall mental map. Also, the Compact Orthogonal Layout algorithm in earlier yFiles~\cite{yWorks} versions explicitly attempted to fit graphs within a given aspect ratio using a divide and conquer approach. However, the algorithm was eventually removed, suggesting that the heuristic-based compactness control did not generalize well across diverse graph structures or application contexts.

Across graph layouts research, there is a clear recognition that controlling the overall aspect ratio of the layout is often desirable, yet it is underexplored in orthogonal graph drawing. Our work introduces aspect ratio as a soft constraint within the orthogonal layout stress-based optimization process, enabling the generated layouts to automatically adapt to arbitrary target proportions (e.g., 16:9, 4:3, or 1:1) while maintaining the quality of the layout.

\subsection{Force-Directed Layout}

Classical force-directed methods do not strictly enforce aspect ratio; they often let the layout fill whatever shape emerges, though implementations sometimes confine the nodes to an initial square or use cooling strategies to keep the graph roughly square. Koren's work~\cite{Koren2005} on drawing graphs with eigenvectors includes constraints to force the nodes to be equally scattered along each of the axes, resulting in a perfectly balanced aspect ratio in the final graph.  Some force-directed variants add custom forces or constraints to shape the layout. For example, ImPrEd, an improved force-directed algorithm by Simonetto \etal~\cite{Simonetto2011} includes modifications that help maintain a good aspect ratio of the graph. Modern force-directed graph libraries expose these controls, e.g. the yWorks Organic~\cite{yWorks} layout provides an output-restriction option, so that the layout result can be specified to fit into a desired aspect ratio. A more recent approach by Di Giacomo \etal~\cite{Giacomo2015} explores dynamic adaptation of aspect ratio to different screen sizes. They adapt image seam carving to networks by identifying and compressing low-importance regions in the layout so that the graph can be resized to a new aspect ratio while preserving node connectivity and overall topology. These approaches can achieve a specified target aspect ratio, but are mostly post-processing techniques and do not integrate the aspect ratio into the layout production process.

A general approach is to constrain some of the node positions to a given region. Zhang and Pang~\cite{Zhang2016} introduced Graph Layout with Versatile Boundary Constraints, which respects user-drawn boundary shapes to specify where some nodes should be located. Users can interactively draw one or more boundaries to spread different density distributions within the defined boundaries. While their method enables user-defined regions in the force-directed layout, it does not enforce a global aspect ratio of the boundary. Li \etal~\cite{Reyan2020} incorporate a differentiable aspect ratio term whose gradients act as a soft scaling force during optimization. Their use of soft boundaries and cross-entropy loss indirectly stretches the graph in the $x$ and $y$ dimensions until the layout becomes approximately square. This mechanism is conceptually related to our normalization step, as both approaches influence aspect ratio during the iterative force-directed stage. However, their scaling is implicit with multiple loss terms, whereas our method applies an explicit geometric normalization based on variance ratios after the computation of the forces. Also, it is hard to approximate the area of the graph in our case, since not all the nodes of the graph are involved during stress minimization.


Although force-directed algorithms that account for aspect ratio constraints have been developed,
these techniques are rarely applied to orthogonal layouts. Our method fills this gap by embedding a soft aspect ratio constraint into the HOLA framework. It achieves screen-fitting orthogonal layouts with low added effort. We use a normalization step, a geometric scaling of node coordinates around their centroid~\cite{Lucas2023,Jin2004,WANG2012,Raschka2014}, that continuously adjusts the node distribution to approach a target aspect ratio.



\begin{figure*}[t]
  \centering
  \includegraphics[width=\textwidth,trim={0 3cm 0 0.4cm},clip]{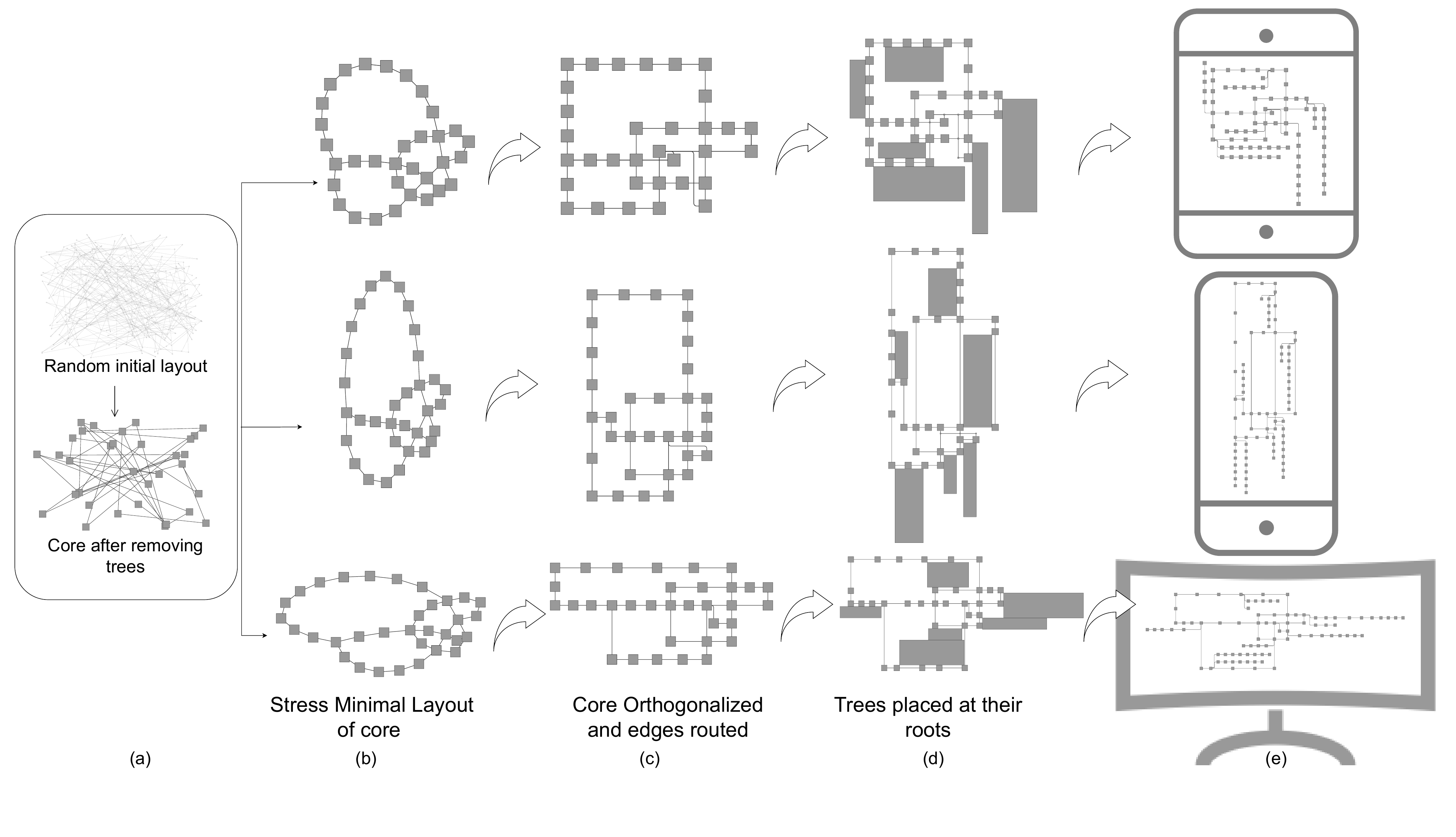}
  \vspace{-2mm}
\caption{ The main stages of ARCOL pipeline for the same graph for three different aspect ratios. (a) From random initial layout to core without trees. (b) stress minimization with the normalization step to guide the layout to the shape that matches the target aspect ratio. (c) core orthogonalization. (d) trees attachment (e) final graph fitted to the target aspect ratio. }
  \label{fig:steps}
  \vspace{-3mm}
\end{figure*}

\section{Method}
In this section we explain the HOLA pipeline and detail our aspect ratio constraint integration. Figure~\ref{fig:steps} illustrates the individual steps of the pipeline, demonstrating how the aspect ratio is enforced at multiple stages to produce a layout that fits a target AR. 

\subsection{Preliminaries: Revisiting HOLA}
HOLA~\cite{Kieffer2016} is an automatic layout algorithm designed to produce orthogonal network layouts that resemble those made by humans. HOLA replaces the classical Topology-Shape-Metrics pipeline with a sequence of incremental, human inspired steps guided by three principles: (P1)~use stress minimization to reveal symmetry and reduce edge length variance; (P2)~apply local, opportunistic improvements that introduce grid alignments and aesthetic bend points; and (P3)~treat tree-like subgraphs separately so they can be given a symmetrical layout and placed aesthetically. The algorithm first decomposes the input graph into a biconnected core and a set of peripheral trees, computes a stress-minimizing
layout for the core, orthogonalizes it, then reinserts the trees into the faces of the planarized core using greedy placement heuristics.
A final refinement stage performs local alignment and node distribution.

While HOLA produces visually balanced and compact layouts, it does not explicitly control the aspect ratio of the final graph. Instead, the final aspect ratio is an unpredictable consequence of the graph's structure and the initial node positions. To address this limitation, we apply a normalization method in HOLA's stress minimization, and a custom cost function in its tree reattachment phase, in order to guide the layout toward the desired aspect ratio.

\subsection{Normalized Stress Minimization}

We introduce a simple variance-based normalization step into each iteration of the stress minimization phase. This normalization measures the spread of node positions along each axis using coordinate variances. In this section, we explain the method in detail and describe the normalization procedure used to scale the overall shape of the layout toward a predefined aspect ratio. After each iteration of the stress minimization loop, the current spatial spread of node positions along the horizontal and vertical axes is measured and rescaled. The normalization gradually adjusts the layout toward the target aspect ratio, while preserving the relative spatial relationships established by the force-directed optimization. The normalization method is inspired by covariance normalization in
multivariate statistics~\cite{Duthie2024,boos2004}, where the variance along each dimension describes the spread of data points. In our layout context, the variances of the node coordinates along the $X$ and $Y$ axes provide a proxy for the graph's width and height. Then the square root of the variance acts as a proxy for the aspect ratio, which is more robust to outlier nodes compared to using ($X_\text{max} - X_\text{min})/(Y_\text{max} - Y_\text{min})$ of the coordinates directly. Also, because the nodes are not final, and this step is only applied on the core without the trees, a more stable measure is beneficial over an exact measure of the incomplete graph. By rescaling these variances to a target ratio, we can control the global aspect ratio without introducing discontinuities or disturbing the internal geometry.

HOLA's layout of the core graph begins with an initial ``distribution'' phase, in which a first pass of unconstrained stress-minimization~\cite{DwyerKim2006}, and second pass with overlap removal constraints~\cite{DwyerPeter2005} establish a basic shape for the core. By applying our normalization during these steps, we are already able to draw the shape of the core toward the target aspect ratio (Figure~\ref{fig:steps}(b)), before HOLA's orthogonalization (Figure~\ref{fig:steps}(c)) has begun. As HOLA then proceeds through its later phases, it repeatedly uses stress minimization to dissipate stress accumulated by the application of constraints. In these later applications, it is important that we continue to apply our normalization, so that the aspect ratio can be maintained, and even improved.


Let $\bar{X}$ and $\bar{Y}$ denote the centroids of all nodes' coordinates along each axis, and let
$\operatorname{var}(X)$ and $\operatorname{var}(Y)$ be the variances of
their coordinates along each axis.
From these, we use the ratio of standard deviations as a proxy for the current aspect ratio of the layout,
\[\text{AR}_{\text{cur}} = \frac{\sigma X}{\sigma Y} = \sqrt{\frac{\mathrm{var(X)}}{\mathrm{var(Y)}}}\]  
We then compute scaling factors that steer the layout toward the target aspect ratio $AR^*$ while preserving its
area:
\[
s_x = \sqrt[4]{\frac{\text{AR}^*}{\text{AR}_{\text{cur}}}}, \qquad
s_y = \frac{1}{s_x}
\]
The scaling is applied around the centroid so that the layout remains
centered: $X_i' = \bar{X} + s_x (X_i - \bar{X})$ and
$Y_i' = \bar{Y} + s_y (Y_i - \bar{Y}),$
where $X_i', Y_i'$ are the new coordinates of node $i$. This transformation expands one axis, compresses the other, and preserves the overall area of the layout while gradually
bringing its aspect ratio closer to the target. These steps are outlined in Algorithm~\ref{alg1}.

A natural choice for $s_x$ is the square root of the ratio of ARs, since this would achieve a full correction.
We instead use the fourth root, applying only half that correction, because when the layout is far from the target aspect ratio, the square root can be too far from 1, causing large jumps in node positions. Applying this correction fully after each iteration often leads to oscillation or overshooting, since the aspect ratio fluctuates during stress minimization.
The fourth root softens these corrections by reducing their magnitude when the AR error is large while making almost no difference when the layout is already close to the target. This way we stabilize the iterative optimization and prevent overshooting.


Since the stress-minimization process can converge to local minima, we perform several rounds of HOLA's distribution phase using our normalization, each starting with shuffled node positions. Among these runs, the layout coming closest to the target AR (and with lowest stress, to break ties) is selected as the initial configuration shown in Figure~\ref{fig:steps}(b). After this, orthogonalization on the core proceeds. We did not alter HOLA's rules for orthogonalization; its repeated use of stress-minimization during this phase, modified by our normalization, was enough to maintain the desired aspect ratio, as can be seen in Figure~\ref{fig:steps}(c). We observed that the aspect ratio does not change much during this orthogonalization step.


\begin{figure*}[t]
\centering
\includegraphics[width=\textwidth]{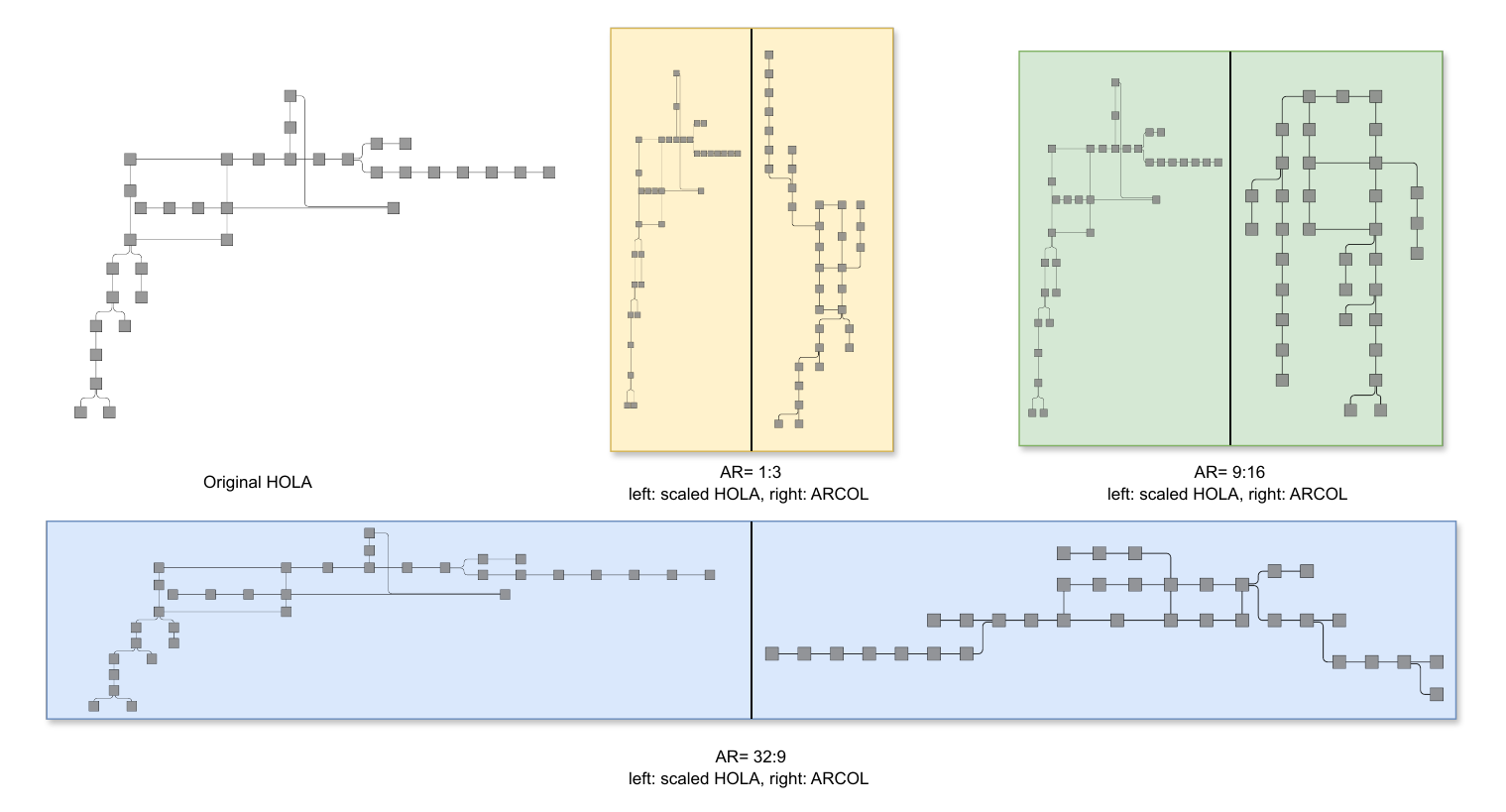}
\vspace{-3mm}
 \caption{Comparison of post-scaled HOLA and ARCOL for a graph from 
 Rome dataset with different aspect ratios.}
 \label{grafoFigure}
 \vspace{-3mm}
\end{figure*}

\begin{figure*}[h]
\centering
\includegraphics[width=\textwidth]{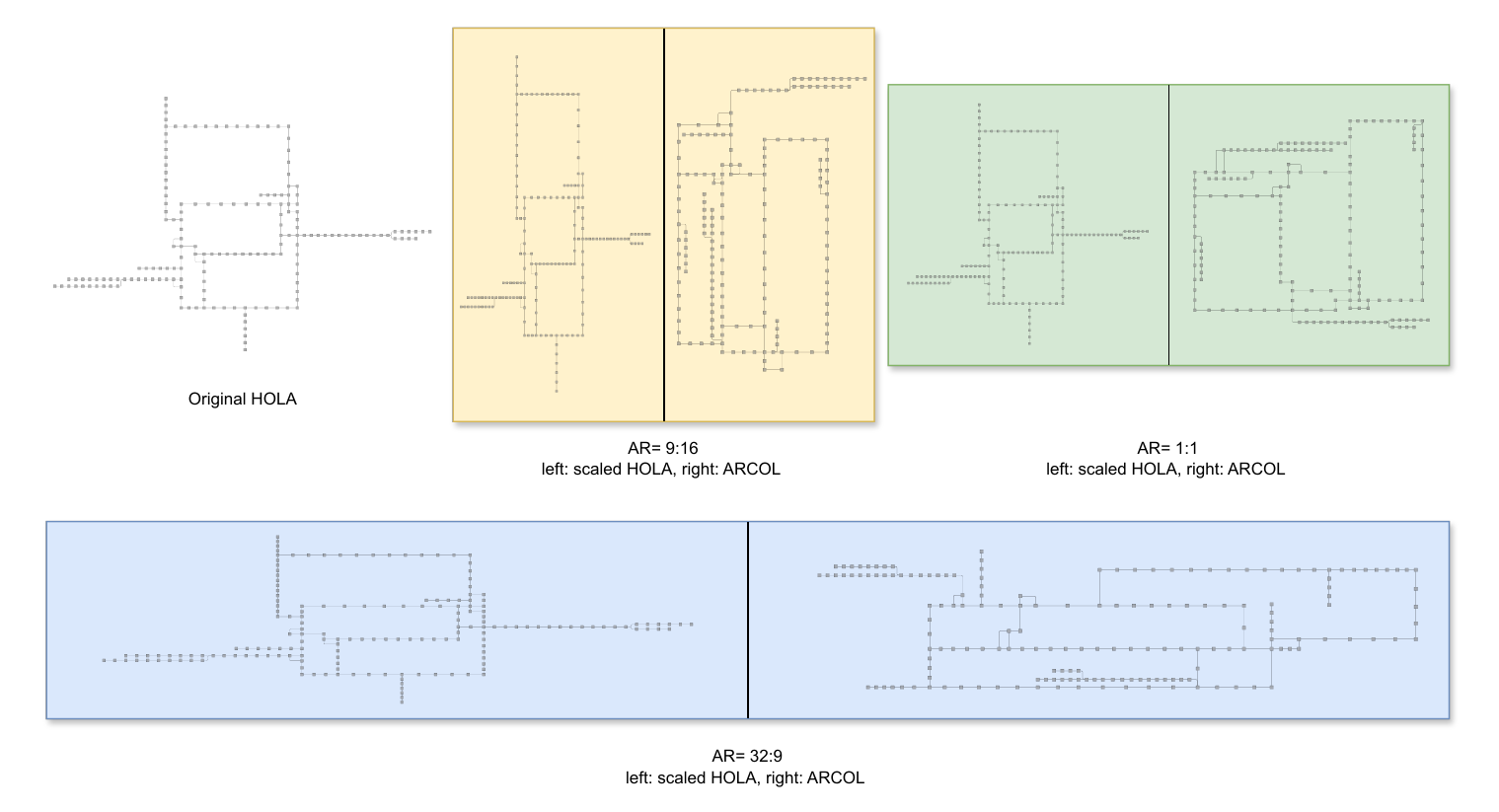}
\vspace{-3mm}
 \caption{Comparison of post-scaled HOLA and ARCOL for Sydney metro map graph with different aspect ratios.}
\label{SydneyFigure}
\vspace{-3mm}
\end{figure*}   

\subsection{Tree attachments}
During the reattachment phase of the pipeline, peeled trees must be placed onto faces of the planarized core graph in a way that preserves readability while steering the global graph toward a user-specified target aspect ratio. We formulate this as a cost-only optimization problem that evaluates every possible tree placement by a unified cost function combining geometric expansion cost and aspect ratio deviation.
The placement with the lowest total cost is selected greedily, proceeding from the largest trees to the smallest.

In the original formulation of HOLA, trees have their symmetric layout separately and are considered greedily based on their bounding boxes' areas. Then, tree placement was guided primarily by geometric heuristics such as favoring cardinal directions and external faces, and secondarily by minimizing local expansion cost. In contrast, our method introduces a unified cost model that directly incorporates aspect ratio considerations into the placement decision.

\begin{algorithm}
\caption{Aspect Ratio--Aware Stress Minimization}
\For{iteration in stress\_minimization}{
    compute forces and update positions.\\
    compute variance ratio $\text{AR}_{\text{cur}} = \sqrt{\frac{\mathrm{varX}}{\mathrm{varY}}}$\\
    $s_x = \sqrt[4]{\frac{\text{AR}^{*}}{\text{AR}_{\text{cur}}}} $ ,
    $s_y = \frac{1}{s_x} $\\
    rescale coordinates around centroid.\\
    compute stress and check convergence.\
}
\label{alg1}
\end{algorithm} 

For each candidate placement defined by a \emph{target face}, an \emph{orientation}, and a \emph{flip bit}, we first estimate the horizontal and vertical expansion costs required to fit the tree into the chosen face.
These costs, $C_X$ and $C_Y$, are computed by the existing HOLA expansion model that accounts for available space, separation constraints, and node padding. We then predict the projected bounding box of the layout if the tree were inserted
at this position and derive its projected aspect ratio $\text{AR}_{\text{proj}} = \frac{W'}{H'},$
where $W'$ and $H'$ denote the resulting layout width and height, respectively. 




Our cost function,
\[
     C_{\text{final}} = C_{\text{space}} + \lambda \, \omega \, C_{\text{AR}}
\]
consists of two terms: one that measures the way the tree placement uses space locally, and a second term that measures the way it will affect the aspect ratio globally. The impact of the second term is scaled by two weighting factors, $\lambda$ and $\omega$, defined below.

The space cost is a weighted combination of the costs $C_X$ and $C_Y$ of expanding in each dimension,
\[
    C_{\text{space}} = w_X C_X + w_Y C_Y
\]
where the weights,
\[
(w_X, w_Y) =
\begin{cases}
(0.85, 1.0), & \text{if } \text{AR}_{\text{proj}} < \text{AR}^* \text{ (layout too tall)},\\[3pt]
(1.0, 0.85), & \text{if } \text{AR}_{\text{proj}} > \text{AR}^* \text{ (layout too wide)},\\[3pt]
(1.0, 1.0),  & \text{otherwise.}
\end{cases}
\]
are designed with a mild $15\%$ discount that favors width- or height-increasing placements when they improve the aspect ratio.



To explicitly penalize the deviation from the target aspect ratio, we add a smooth, symmetric
\emph{log-squared penalty}:
\[
C_{\text{AR}} = \log^2\!\left(\frac{\text{AR}_{\text{proj}}}{\text{AR}^*}\right),
\]
which treats over- and under-stretching equally and scales naturally with relative deviation.
The influence of this term is adjusted by a \emph{leverage factor}:
\[
\lambda = \min\!\left(1,\left(\frac{\text{A}_{\text{tree}}}{\text{A}_{\text{core}}}\right)^{\beta}\right),
\]
where $\text{A}_{\text{tree}}$ is the area of the bounding box of the tree
before reattaching it, and $\text{A}_{\text{core}}$ is that of the current bounding box of the core (including any trees that have already been reattached).
This way, large trees can meaningfully adjust the global aspect ratio,
while small trees exert proportionally weaker influence. The exponent $\beta$ is empirically chosen to be 0.75 to control how much the size of each tree affects its ability to change the layout's aspect ratio.
If the value were 1.0, large trees would have too much influence and could distort the layout too quickly.
If it were 0.5, small trees would have too little influence and the layout would change too slowly.
Using $\beta=0.75$ gives a middle ground: big trees can still guide the layout, but small trees also have a small effect.
This adjusts the aspect ratio in a smooth and stable way without sudden jumps or over-corrections.

Finally, to normalize the magnitudes of the spatial and aspect ratio terms,
we multiply the $\lambda C_{\text{AR}}$ penalty by a constant weight
$\omega$ proportional to the \emph{ideal edge length} (as computed by HOLA)
of the current layout.



Each tree is placed into the face and orientation that minimize $C_{\text{final}}$.
This unified cost provides a smooth trade-off between geometric compactness and aspect ratio without introducing hard constraints. Trees are processed in descending order of size to allow larger trees to establish the global aspect ratio early.
For each tree, all candidate placements are enumerated and scored using the above cost.
The minimum cost placement is selected, the corresponding expansion is applied, and the tree is inserted into the layout.
After all trees have been placed, a compaction and routing phase resolves local overlaps and finalizes edge geometry.
Although later refinements may slightly alter the geometry, the placement phase determines the final aspect ratio.

In the final step of the pipeline, the layout is gently rescaled around its center so that its overall width-to-height shape matches the target aspect ratio. The adjustment is again variance-based and limited to small, bounded corrections to apply only minor changes to improve the aspect ratio without distorting the layout’s structure or proportions. 
We first measure the current aspect ratio of the final layout, and if this already lies within $15\%$ of the target, then no correction is applied, avoiding unnecessary distortions. Otherwise, we compute the scaling factors for the X and Y axes:
$s_x = \sqrt{\frac{\text{AR}^{*}}{\text{AR}_{\text{cur}}}}$, $s_y = \frac{1}{s_x} $. To avoid sudden or extreme deformations, we do not apply the full correction. Instead, only $30\%$ of the required change is applied and the scaling is limited to a maximum of $\pm~20\%$ and is always performed around the center of the layout. This final adjustment can also be used to force the graph into a specific aspect ratio, beyond the proposed maximum scaling limit, if desired.


\begin{table}[thb]
    \centering
    \resizebox{\linewidth}{!}{
    \begin{tabular}{clccccccc}
\toprule
AR & Method & $\Leftrightarrow$ 
& KSM$\uparrow$ & ELD$\uparrow$ & NR$\uparrow$ & NU$\uparrow$ & NP$\uparrow$ & EC$\uparrow$ \\
\midrule
* & \textcolor{gray}{HOLA} & \textcolor{gray}\xmark & \textcolor{gray}{0.683} & \textcolor{gray}{0.724} & \textcolor{gray}{0.065} & \textcolor{gray}{0.610} & \textcolor{gray}{0.355} & \textcolor{gray}{0.971} \\
\midrule
\multirow{3}{*}{avg}
& ARCOL & \xmark & \textbf{0.661} & \textbf{0.722} & \textbf{0.062} & \textbf{0.618} & \underline{0.322} & \underline{0.968} \\
& ARCOL & \cmark & \underline{0.647} & \underline{0.716} & \underline{0.055} & \textbf{0.618} & 0.321 & \underline{0.968} \\
& HOLA  & \cmark & 0.632 & 0.699 & 0.047 & \underline{0.609} & \textbf{0.341} & \textbf{0.971} \\
\midrule
\multirow{3}{*}{1:3}
& ARCOL & \xmark & \textbf{0.645} & \textbf{0.717} & \textbf{0.053} & \underline{0.599} & \underline{0.320} & \underline{0.968} \\
& ARCOL & \cmark & \underline{0.618} & \underline{0.706} & \underline{0.042} & \underline{0.599} & 0.317 & \underline{0.968} \\
& HOLA  & \cmark & 0.560 & 0.658 & 0.028 & \textbf{0.610} & \textbf{0.321} & \textbf{0.969} \\
\midrule
\multirow{3}{*}{9:16} 
& ARCOL & \xmark & \textbf{0.670} & \textbf{0.725} & \textbf{0.064} & \textbf{0.616} & \underline{0.322} & \underline{0.968} \\
& ARCOL & \cmark & \underline{0.660} & \underline{0.723} & \underline{0.059} & \textbf{0.616} & \underline{0.322} & \underline{0.968} \\
&  HOLA & \cmark &          0.625 &          0.694 & 0.044 & \underline{0.610} & \textbf{0.338} & \textbf{0.971}      \\
\midrule
\multirow{3}{*}{1}
& ARCOL & \xmark & 0.677 & \textbf{0.728} & \textbf{0.071} & \textbf{0.629} & \underline{0.324} & \underline{0.968} \\
& ARCOL & \cmark & \underline{0.678} & \textbf{0.728} & \underline{0.070} & \textbf{0.629} & \underline{0.324} & \underline{0.968} \\
&  HOLA & \cmark & \textbf{0.682} & \underline{0.718} & 0.064 & \underline{0.610} & \textbf{0.354} & \textbf{0.971} \\
\midrule
\multirow{3}{*}{4:3}
& ARCOL & \xmark & \underline{0.675} & \textbf{0.727} & \textbf{0.070} & \textbf{0.629} & \underline{0.325} & \textbf{0.969} \\
& ARCOL & \cmark & 0.673 & \underline{0.726} & \underline{0.067} & \textbf{0.629} & \underline{0.325} & \textbf{0.969} \\
&  HOLA & \cmark & \textbf{0.680} & 0.721  & 0.063 & \underline{0.610} & \textbf{0.355} & \underline{0.968} \\
\midrule
\multirow{3}{*}{16:9}
& ARCOL & \xmark & \textbf{0.669} & \textbf{0.724} & \textbf{0.065} & \textbf{0.626} & \underline{0.321} & \textbf{0.968} \\
& ARCOL & \cmark & \underline{0.659} & \underline{0.721} & \underline{0.060} & \textbf{0.626} & \underline{0.321} & \textbf{0.968} \\
&  HOLA & \cmark & \underline{0.659} &          0.716 & 0.054 & \underline{0.609} & \textbf{0.353} & \textbf{0.968} \\
\midrule
\multirow{3}{*}{21:9}
& ARCOL & \xmark & \textbf{0.659} & \textbf{0.721} & \textbf{0.059} & \textbf{0.619} & \underline{0.320} & \underline{0.967} \\
& ARCOL & \cmark & \underline{0.639} & \underline{0.715} & \underline{0.051} & \textbf{0.619} & \underline{0.320} & \underline{0.967} \\
&  HOLA & \cmark & 0.631             & 0.705 & 0.045 & \underline{0.609} &           \textbf{0.345} & \textbf{0.968} \\
\midrule
\multirow{3}{*}{32:9}
& ARCOL & \xmark & \textbf{0.636} & \textbf{0.711} &\textbf{ 0.050} & \textbf{0.610} & \underline{0.320} & \underline{0.968} \\
& ARCOL & \cmark & \underline{0.606} & \underline{0.696} & \underline{0.037} & \textbf{0.610} & 0.318 & \underline{0.968} \\
&  HOLA & \cmark &             0.591 &               0.681 & 0.031 & \underline{0.608} & \textbf{0.324} & \textbf{0.970} \\
\bottomrule
    \end{tabular}}
    \caption{\textbf{Graph Metrics} comparing ARCOL to the HOLA baseline. All metrics are scaled in $[0,1]$ where higher indicates better. $\Leftrightarrow$ indicates stretching the result to the target aspect ratio. We highlight the best values per AR in \textbf{bold} and \underline{underline} the second best. \textsuperscript{*}HOLA at the aspect ratio it produces, typically close to 1:1.}
    \label{tab:metrics}
\end{table}

\section{Results}
In this section, we show some of the results of ARCOL on two graphs with different aspect ratios and compare it to the baseline HOLA and to post-scaled HOLA. Across all tested graphs from the Rome dataset and target aspect ratios, ARCOL produces layouts that match the desired aspect ratio while preserving the structural clarity of HOLA orthogonal drawings quality metrics. 
Figures~\ref{grafoFigure} and \ref{SydneyFigure} illustrate typical outcomes.
ARCOL creates layouts that fit the desired shape while still preserving grid alignment and local symmetry. In contrast, post-scaled HOLA  simply stretches the layout which results in distorted edge lengths, unbalanced spacing, and regions that appear compressed or overly elongated. We show more of the qualitative results in the supplementary material.

\section{Evaluation}
To evaluate the visual quality and readability of our aspect ratio-aware layouts, we evaluate our layouts along multiple axes. First, we adopt the universal graph-drawing metrics proposed by Mooney \etal~\cite{Mooney2025}. These metrics were designed to provide unambiguous, normalized measures of key aesthetic principles such as edge crossings, edge length deviation and stress over both straight-line and polyline drawings. Their definitions were recently extended to handle curved or orthogonal edges and are implemented in the open source GEG framework (\httpAddr{github.com/gavjmooney/geg}). Second, we perform a user study with 72 participants. In the study, we compare our method to HOLA and ask the participants to rate the readability of the layouts, how well they adhere to a target aspect ratio and which they prefer overall.
Lastly, we conduct an interview with three layout experts.
In order to compare to HOLA as a baseline for a specific aspect ratio, we use an area preserving re-scaling of the graph produced by HOLA to achieve the target aspect ratio. Figures~\ref{grafoFigure} and~\ref{SydneyFigure} show some results and comparisons between post scaled HOLA and ARCOL. 

\begin{figure}[thb]
    \centering
    \includegraphics[width=\linewidth]{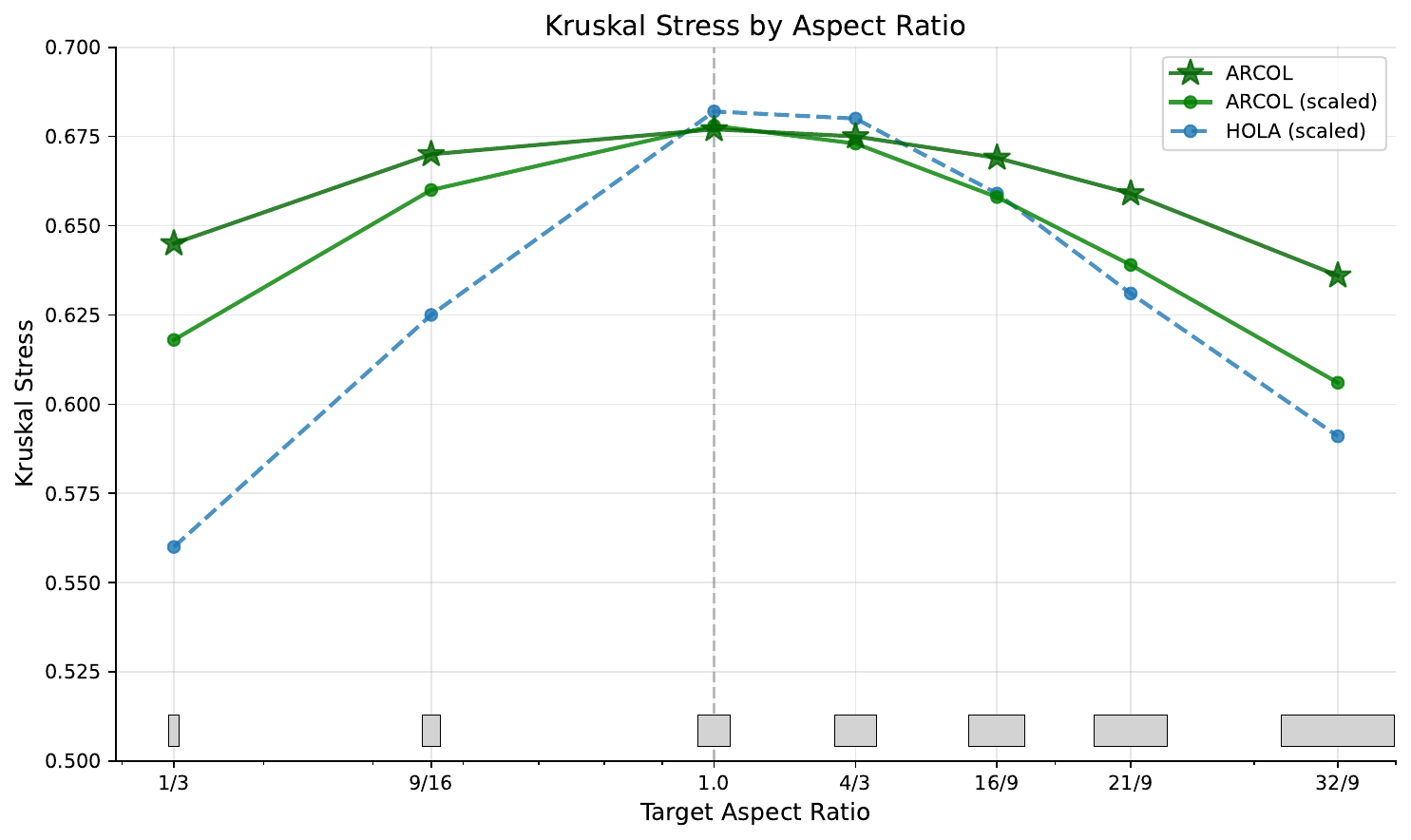}\\
    \includegraphics[width=\linewidth]{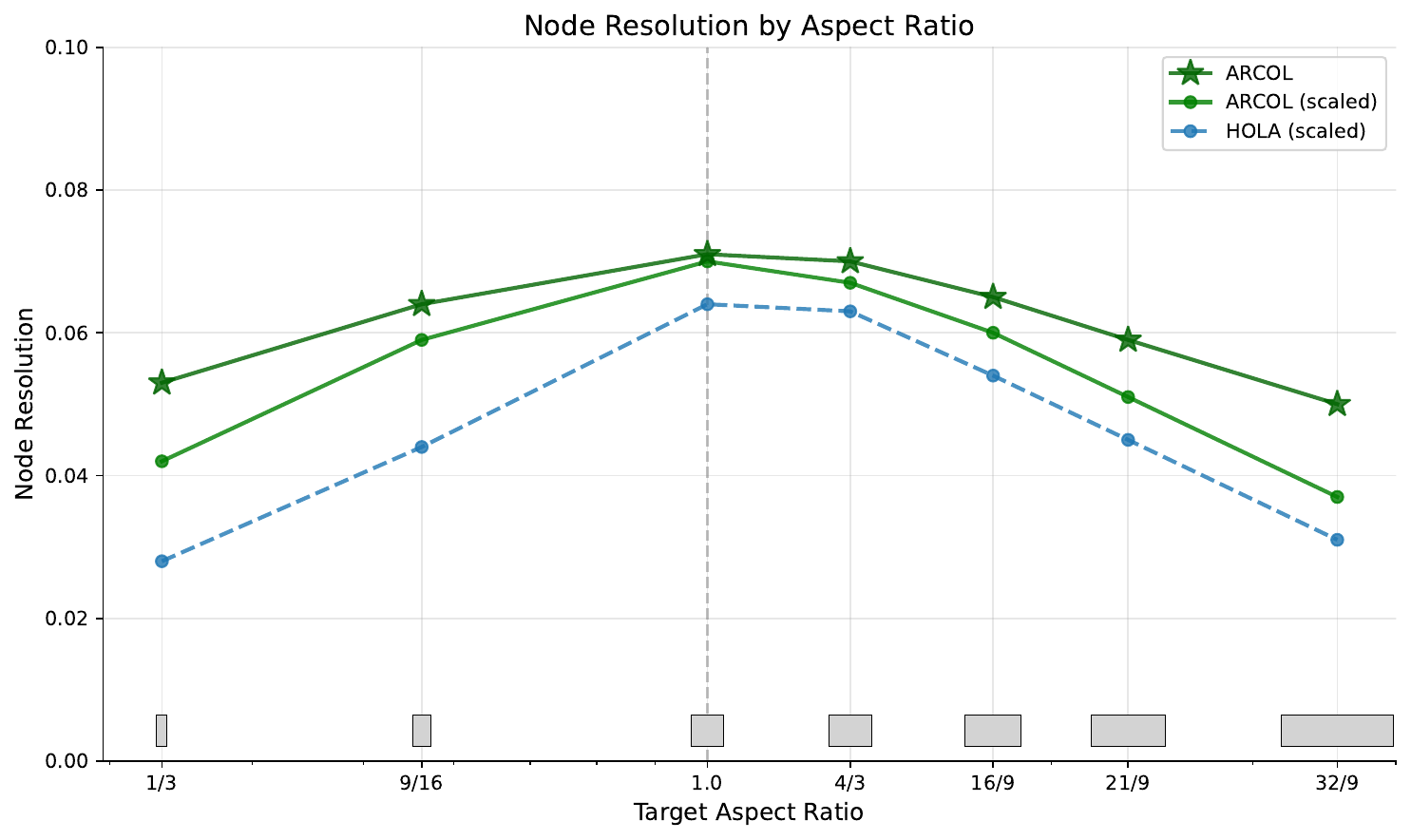}
    \caption{Comparison of Kruskal Stress and Node Resolution between ARCOL and HOLA. (Higher scores are better.)}
    \label{fig:metrics-comparison}
\end{figure}

\subsection{Graph Metrics}
To assess the quality of our layouts compared to HOLA, we choose the following set of metrics, focusing on the most relevant indicators for orthogonal layouts and the aspect ratio constraint (see~\cite{Mooney2025} for more details).


\noindent\textbf{Kruskal Stress (KSM)} measures how well the Euclidean distances between nodes reflect their relative ordering distances. Higher values correspond to better structural preservation.

\noindent\textbf{Edge-Length Deviation (ELD)} is the standard deviation of edge lengths relative to the mean. Higher values imply a more uniform and visually balanced edge distribution.

\noindent\textbf{Node Resolution (NR)} is the ratio between the smallest and largest pairwise node distances. A higher value implies clearer visual separation between vertices and fewer overlaps.

\noindent\textbf{Node Uniformity (NU)} quantifies how evenly nodes are distributed across the drawing area, contributing to visual balance and perceptual uniformity.

\noindent\textbf{Neighborhood Preservation (NP)} measures how well the spatial neighborhoods in the layout correspond to topological adjacency in the graph. Higher values indicate stronger local structure preservation.

\noindent\textbf{Edge Crossings (EC)} is the normalized number of intersecting edge pairs. Fewer crossings generally improve readability and reduce visual clutter, and fewer crossings are indicated by higher EC values in the GEG framework.

Together, these metrics evaluate complementary aspects of layout quality, from geometric precision and structural faithfulness to
spatial uniformity, allowing a comprehensive quantitative comparison between our method and baseline HOLA layouts. For each layout, we also
record the number of nodes, edges, and maximum node degree to normalize and interpret metric values relative to graph size and density. We use 1481 graphs randomly chosen from the Rome dataset~\cite{DiBattista1997} and tested with 7 different aspect ratios.

Table~\ref{tab:metrics} reports our findings with averages of each metric overall (AR=avg), as well as for specific aspect ratios. To assess whether the aspect ratio control meaningfully degrades the metrics, we also report results from HOLA without any AR modification (first row) to get a baseline for the chosen metrics. Comparing standard HOLA at its favored AR to ARCOL on average over various aspect ratios (row 1 vs. 2), reveals that our aspect ratio constraint incurs a steady, but mostly very minor reduction in the metrics, while even improving on the node uniformity metric.

\begin{figure}[tbh]
    \centering
    \includegraphics[width=\linewidth]{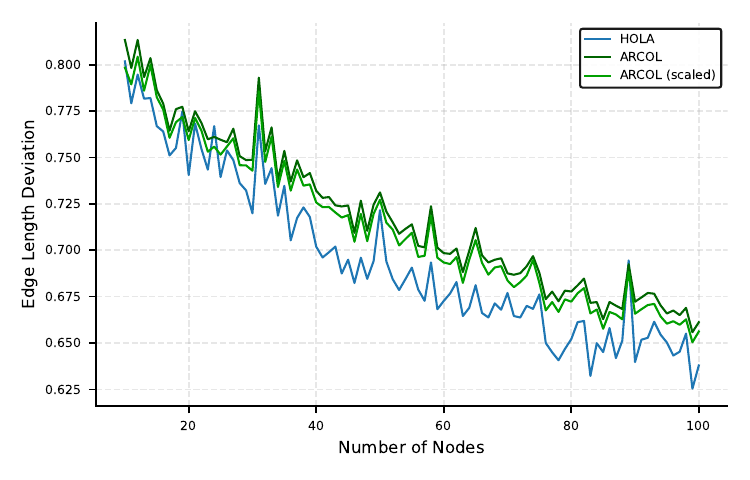}
    \vspace{-6mm}
    \caption{Comparison of ELD over different graph sizes.}
    \label{fig:metrics-by-size}
    
\end{figure}

Comparing ARCOL to scaled HOLA for different aspect ratios reveals that for square-ish AR's, HOLA achieves slightly lower stress, while already suffering for metrics like edge length deviation, node resolution and node uniformity, probably because even slight re-scaling can reduce these metrics. For more extreme AR's, i.e. tall or wide layouts, we see that ARCOL achieves a meaningful improvement over scaled HOLA in stress, edge length deviation, node resolution and uniformity. This effect is more clearly illustrated in Figure~\ref{fig:metrics-comparison}. 
Since we found graph size to also have a large influence on resulting layouts, we further compare edge length deviation over different graph sizes in Figure~\ref{fig:metrics-by-size}. We present plots for the remaining metrics split by aspect ratio and over graph size in the supplementary material.

Lastly, we find that edge crossings are not affected by any re-scaling and are the result of the initial layout. Since ARCOL adds the soft AR constraint during stress minimization, we achieve slightly different layouts which typically lead to a negligible downside in edge crossings. Similarly, this constraint pushes apart some neighbors in favor of achieving the target aspect ratio, leading to slight reduction in neighborhood preservation compared to HOLA.

\subsection{User Study}
We conducted an online user study on the Prolific (\httpAddr{//www.prolific.com/}) website with 72 participants. We used 33 graphs and 6 different aspect ratios. Participants were asked to compare ARCOL to scaled HOLA layouts (comparable to Fig.~\ref{grafoFigure} and \ref{SydneyFigure}). then, participants were asked to express which layout they like better in terms of clarity, balance, edge uniformity, efficiency of space use and overall preference on a 5-point Likert scale. Each comparison received ratings from at least 5 different judges. 

For evaluation, we normalized these Likert scales to a -2 (strong HOLA preference) to +2 (strong ARCOL preference) scale. 
A one-sample t-test revealed that the mean preference rating ($M = 0.17$, 
$SD = 1.66$) was significantly different from zero, 
$t(1475) = 3.89, p < 0.001$ (one-tailed), 
95\% CI $[0.08, 0.25]$, indicating a preference for ARCOL over HOLA. The effect size was 
negligible (Cohens $d = 0.10$).
Given potential non-normality in the data, we confirmed this finding using 
a non-parametric Wilcoxon signed-rank test ($W = 471778, p < 0.001$), 
which yielded consistent results.
Of the 1476 ratings, 724 (49.1\%) 
indicated a preference for ARCOL, 579 (39.2\%) 
indicated a preference for HOLA, and 173 (11.7\%) 
showed no preference.

\begin{figure}[tbh]
    \centering
    \includegraphics[width=\linewidth]{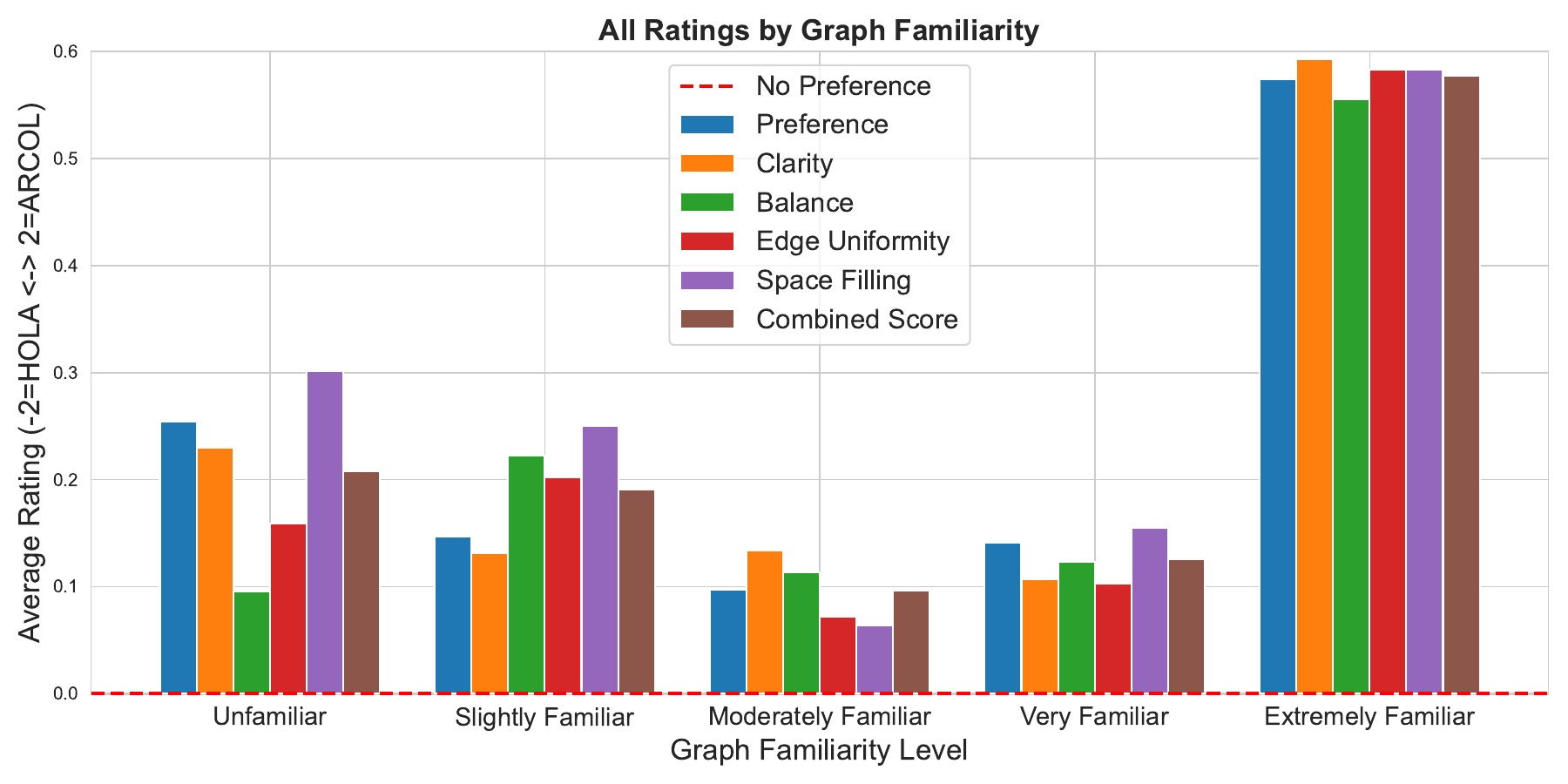}
    \vspace{-6mm}
    \caption{User Study Ratings by Expertise}
    \label{fig:userstudy-familiarity}
\end{figure}

\begin{figure}[tbh]
    \centering
    \includegraphics[width=\linewidth]{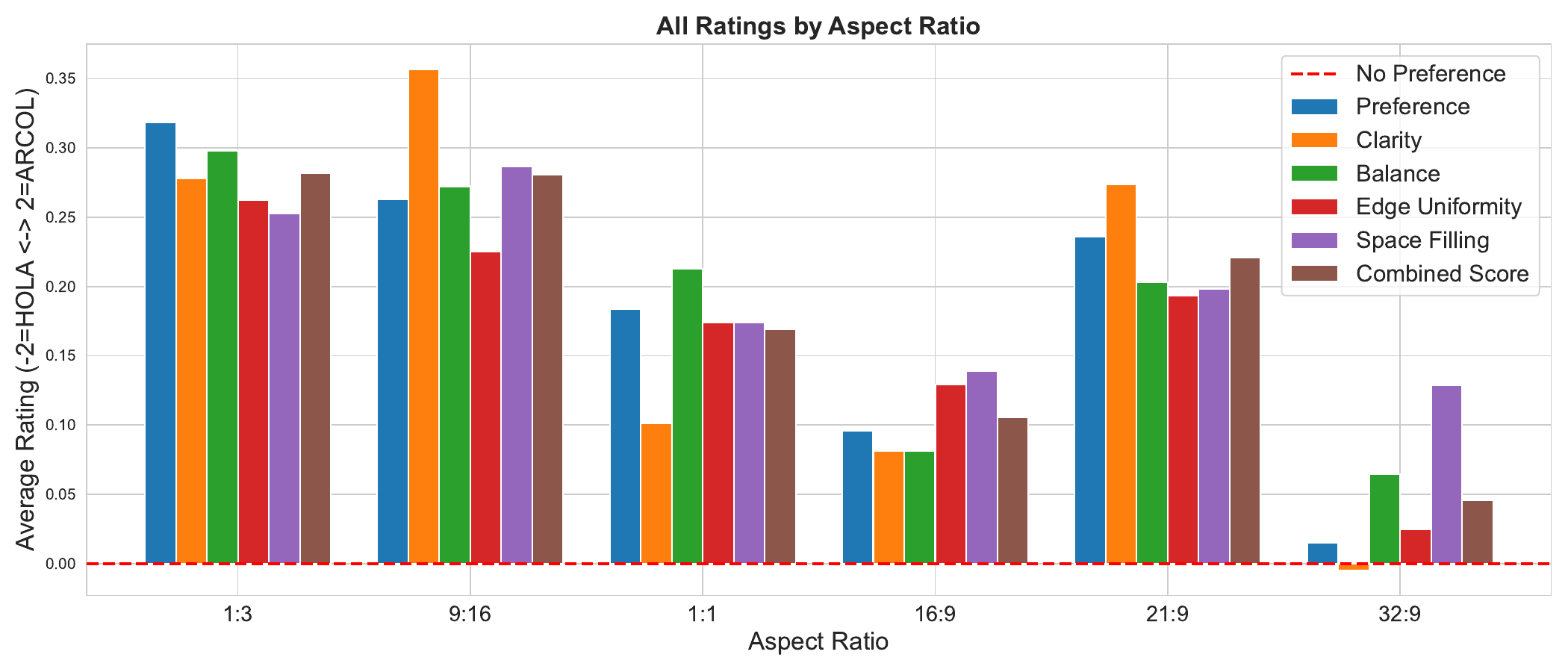}
    \vspace{-6mm}
    \caption{User Study Ratings by Aspect Ratio}
    \label{fig:userstudy-aspectratio}
\end{figure}

Furthermore, we found that participants that reported high familiarity with orthogonal layouts have a much stronger preference for ARCOL than inexperienced viewers. The average ratings grouped by experience are illustrated in Figure~\ref{fig:userstudy-familiarity}. We also looked at the rankings grouped by the different aspect ratios. The results for this are shown in Figure~\ref{fig:userstudy-aspectratio} and show that ARCOL is especially preferred with tall layouts. Overall the preference for ARCOL is consistent, However, once very extreme aspect ratios such as 32:9 are reached, the preference becomes less strong. This may partly reflect the specific graphs shown at these ratios or the fact that HOLA’s baseline style is naturally more horizontal, which can make very wide shapes feel more familiar to some viewers.

\subsection{Expert Interviews}
\label{sec:interviews}
We interviewed three experts from yWorks, each with nearly two decades of experience in graph layouts and algorithms, to obtain feedback on our proposed approach. Across all interviews, the experts reported extensive hands-on experience with orthogonal graph drawing. Two experts specifically described work on traditional orthogonal layout algorithms, layered/Sugiyama-style approaches, orthogonal edge routing, and TSM (Topology-Shape-Metrics) frameworks. All three experts confirmed that explicit aspect ratio constraints are not part of existing orthogonal layout pipelines. When aspect ratio is considered at all, it is handled only as post-processing through scaling. The experts independently highlighted recurring issues with orthogonal layouts:
\begin{itemize}
    \item drawings often grow too wide or tall, due to the graph structure;
    \item users often need to resize diagrams to make them fit;
    \item grid alignment and separation constraints make post-processing highly sensitive, as small adjustments can introduce more bends;
    \item compactness competes with readability, and adjusting one dimension affects bend count, density, and whitespace distribution;
    \item with TSM-based approaches, certain areas may become overly constrained, creating unnatural whitespace or tight packing.
\end{itemize}

All experts emphasized that aspect ratio matters whenever diagrams must fit predefined regions, including UI containers and dashboards, multi-panel environments, printed documents, reports and export formats, and screens with fixed width and height.
One expert stressed that the numerical value of the aspect ratio is less important to users than simply ensuring that: “The most important is that it fits in the screen.”

They confirmed numerous real-world problems arising from AR mismatch: horizontally or vertically overflowing diagrams, wasted screen space, overly dense areas or large empty regions, manual stretching or shrinking, and frequent cropping when the layout does not match the shape of the display. They described this as a long-standing issue in industrial tools such as UML editors, workflow visualizations, and technical schematics. None of the experts were aware of any orthogonal layout method that directly optimizes aspect ratio. Overall, the experts’ evaluations of our results were consistently positive. They highlighted that, in several cases, ARCOL achieves a more compact and space-efficient layout than HOLA, which they considered a valuable contribution to improving the clarity and practicality of orthogonal graph drawings. They noted that orthogonal style drawings can often absorb distortion more gracefully than straight-line layouts. They also observed that some graphs naturally adapt better to tall or wide shapes due to their intrinsic structure, and that expecting identical structure across all AR conditions is unrealistic. All experts agreed explicitly or implicitly that our method is genuinely novel, because it introduces aspect ratio awareness inside the layout optimization rather than applying it afterward.
One expert summarized the gap clearly: “Aspect ratio is a real functional constraint, but layout algorithms have not treated it as a first-class target.” Another described the contribution as: “A meaningful contribution… no one really solved this inside the orthogonal pipeline.”
The experts considered the trade-offs between AR control and other metrics reasonable, noting that slight distortions or redistributions of space are acceptable, given the benefit of predictable AR behavior. All three experts stated that this functionality could be integrated into production tools like yFiles, where users work with fixed display regions. They suggested practical improvements related to achieving even more compactness; however, they agreed that AR control addresses a real and unmet need, solving what one expert called: “A long-standing complaint” about diagrams being too wide or too tall.
\section{Conclusion and Future Work}
In this work, we introduced ARCOL, the first orthogonal layout method that integrates aspect ratio control directly into the layout computation. Existing orthogonal algorithms produce drawings whose width and height are not controlled by the graph structure and optimization pipeline. As confirmed by our expert interviews, AR control is a long-standing practical need but has never been treated as a first-class objective inside orthogonal layout algorithms. We address this gap through two complementary mechanisms: \textbf{(1)} a soft, variance-based normalization during stress minimization that gradually steers node positions toward a target aspect ratio without disrupting structural clarity; and \textbf{(2)} an aspect ratio aware tree placement cost that biases expansions of the core graph in a way that promotes the desired proportion. A limitation of our current pipeline is that trees are removed before layout optimization and only reattached after the core graph has reached its target aspect ratio. Integrating the trees earlier in the process could potentially provide additional flexibility, as their bounding boxes might explore different orientations or placements during the stress-minimization phase, which is a challenging combinatorial and geometric problem on its own. 

In addition to the ARCOL algorithm itself, we make three surprising observations:
\textbf{(A)} Injecting continuous, gentle AR adjustments into the stress minimization loops does not lead to instability. The stress minimization process does not ``snap'' the layout back to a square aspect ratio, but merely minimizes stress for the layout induced by AR adjustments. \textbf{(B)} These continuous AR adjustments produce better results if they are based on a proxy for AR that is more stable than the actual AR. \textbf{(C)} In our evaluations, for any target AR, these continuous AR adjustments eventually lead to a stable layout whose actual AR is typically close to the target AR. This suggests that stress minimization techniques effectively explore a space that exhibits many local minima, and that our AR adjustment steers the algorithm toward those minima. This phenomenon might not be specific to AR adjustments, and perhaps introducing gentle steering based on other types of soft constraints could reveal similar patterns and local minima. If so, our approach might be generalizable to different kinds of constraints.

There are several promising directions for future work. A natural next step is to explore joint optimization of compactness, bend minimization, and aspect ratio, enabling users to define custom trade-offs tailored to their application needs. Another is to extend ARCOL toward interactive and dynamic scenarios, where the layout adapts in real time as the viewing window changes. Finally, broadening the evaluation to include task-based user studies, measuring speed and accuracy on graph reading tasks, would provide deeper insight into the practical benefits.

Our quantitative evaluation demonstrates that ARCOL achieves balanced, space-efficient layouts across all tested aspect ratios and outperforms post-scaled HOLA in extreme AR settings, particularly in stress, edge-length deviation, node resolution, and uniformity. The user study confirms that users perceive ARCOL diagrams as more coherent and better matched to the target AR, and expert interviews validate the novelty, usefulness, and practicality of integrating AR constraints directly inside the orthogonal pipeline.

\section*{Acknowledgments}
This work was supported by King Abdullah University of Science and Technology (KAUST), Saudi Arabia with collaboration with TU-Wien, Austria. We thank the yWorks experts who participated in our interviews and provided valuable feedback on the design and evaluation of our approach.

\bibliographystyle{eg-alpha-doi} 
\bibliography{paper}       


\newpage

\end{document}